\documentclass[a4paper,preprint]{revtex4-1}
\usepackage{geometry} 
\usepackage{enumerate}
\usepackage{amsmath}
\usepackage{amssymb}
\usepackage[dvips]{graphicx}
\usepackage{color}
\usepackage{tabularx}
\usepackage{gensymb}
\usepackage[footnotesize]{subfigure}
\newcommand{\angstrom}{\textup{\AA}}

\newcommand{\vect}[1]{\textbf{\textit{#1}}}

\begin{document}
\title{Grand-Canonical Adaptive Resolution Centroid Molecular Dynamics: Implementation and Application}
\author{Animesh Agarwal, Luigi Delle Site}
\affiliation{Institute for Mathematics, Freie Universit\"at Berlin, Germany}
\email{luigi.dellesite@fu-berlin.de}
\begin{abstract}
We have implemented the Centroid Molecular Dynamics scheme (CMD) into the Grand Canonical-like version of the Adaptive Resolution Simulation Molecular Dynamics (GC-AdResS) method. We have tested the implementation on two different systems, liquid parahydrogen at extreme thermodynamic conditions and liquid water at ambient conditions; the reproduction of structural as well as dynamical results of reference systems are highly satisfactory. The capability of performing GC-AdResS CMD simulations allows for the treatment of a system characterized by some quantum features and open boundaries. This latter characteristic not only is of computational convenience, allowing for equivalent results of much larger and computationally more expensive systems, but also suggests a tool of analysis so far not explored, that is the unambiguous identification of the essential degrees of freedom required for a given property. 
\end{abstract}

\maketitle
\section{introduction}
The path integral (PI) approach is a powerful method that describes the quantum character of spatial delocalization of atoms in space\cite{fey-hibbs}. 
For systems at low temperature the PI description is mandatory in order to capture their essential physical features, however also at room temperature the PI description is relevant for molecular systems composed of light atoms.
In particular, the efficient implementation of PI idea in Molecular Dynamics (MD) turned the PI technique into an accurate computational tool for simulating various molecular systems (see e.g. \cite{tuckerbook}). The critical aspect of PIMD is that it is rather expensive when compared to standard classical MD and thus its employment in simulation studies has been restricted, so far, to small systems and short time scales; it must also be noticed that in recent years more work has been done so that PIMD calculations are simplified and made accessible to researchers equipped with basic standard computational resources \cite{man1, piglet2, taka1, taka2, man3, cont2, cont3}. However there is another way, complementary to the trend cited above, to access properties of a system without the need of having large or simplified PIMD calculations: it consists of embedding a PI system into a reservoir of low computational cost that assures thermodynamic conditions as if the whole system was described at PI resolution. This idea implies that the PI system is an open system and exchanges energy and particles with a reservoir. In this context, open boundary approaches, based on the idea of space-dependent adaptive molecular resolution resolution, have been developed in large numbers in the last years (see e.g. Refs.\cite{jcp1,annurev,ensing, matej, hadress, truh,pande} and references therein). In particular the authors of this paper during the last years have worked on the development of the Adaptive Resolution Simulation (AdResS) approach in its Grand Canonical-like version \cite{prx,jcpchem,njp}. More recently several PIMD approaches have been successfully implemented in  GC-AdResS \cite{jcppi}, however, one particular approach, that is the Centroid Molecular Dynamics, deserved a more careful testing and implementation. In fact the theoretical and computational complexity of both methods, GC-AdResS and CMD, is such that an efficient and accurate merging of the two in a unified scheme is not obvious. In particular the adiabatic hypothesis required in the CMD algorithms has been never tested before in open boundary systems where also the time scale of a possible response of the reservoir (e.g. if the reservoir is too small) may enter into the game. Moreover, compared to approaches such as Ring Polymer MD \cite{rpmd, manrev} (already successfully implemented in GC-AdResS), CMD, in the so called partially adiabatic CMD (PACMD) \cite{hone}, can be computationally cheaper and thus it may represent a more efficient alternative (see note \cite{expl}).
The two systems chosen in this study are liquid parahydrogen at low temperature and liquid water at ambient conditions; they represent ideal tests to validate the theoretical and computational robustness of the resulting method. Liquid parahydrogen has been used in the past to test the robustness of PI approaches; in particular for AdResS, the extreme thermodynamic conditions represent a further challenge to the principle on which adaptive molecular resolution is based. Liquid water instead is of importance in many fields of simulation since water plays a key role in condensed matter systems broadly intended; light atoms like hydrogen atoms and their key role in the bonding network of the liquid make this system an ideal system for testing the GC-AdResS CMD algorithm. 
\subsection{Path Integral Molecular Dynamics}
A classical Hamiltonian of a particle of mass $m$ under the action of a potential $V(x)$: $H=\frac{{\bf p}^{2}}{2m}+V({\bf x})$ can be transformed into a quantized Hamiltonian via the path integral formalism of Feynman \cite{fey-hibbs,tuckerbook}.
The resulting Hamiltonian is formally equivalent to the Hamiltonian of a polymer ring of $P$ beads circularly connected through springs characterized by $\omega_{P}= \frac{\sqrt{P}}{\beta\hbar}$ ($\beta=1/k_{B}T$), a fictitious mass $m^{'}$ and fictitious momenta ${\bf p}$: $H=\sum_{i=1}^{P}\left[ \frac{\bf p_{i}^{2}}{2m^{'}} + \frac{1}{2} m\omega_{P}^{2}({\bf x}_{i}-{\bf x}_{i+1})^{2} + \frac{1}{P} V({\bf x}_{i}) \right]$.
The formalism can be extended to a N-particle Hamiltonian: $H=\sum_{j=1}^{N}\frac{{\bf p}_{j}^{2}}{2m_{j}}+V({\bf x}_{1},....{\bf x}_{N})$ and in case the spin statistics can be neglected the resulting quantized Hamiltonian is: $H=\sum_{i=1}^{P}\left(\sum_{j=1}^{N}\frac{({\bf p}_{j}^{(i)})^{2}}{2m^{'}_{j}}+\sum_{j=1}^{N}\frac{1}{2}m_{j}\omega^{2}_{P}({\bf x}_{j}^{(i)}-{\bf x}_{j}^{(i+1)})^{2}+\frac{1}{P}V({\bf x}_{1}^{(i)},....
{\bf x}_{N}^{(i)})\right)$; it must be noted that the potential acts between beads with same index $i$. The spatial oscillations/fluctuations
of the polymer rings describe, in an effective way, the quantum spatial delocalization of the $N$ atoms. As a consequence the statistical
sampling of the individual bead trajectories, produced (e.g.) by Molecular Dynamics, allows for the calculation of statistical properties of atomic/molecular systems where the quantum effects due to the spatial delocalization of atoms are of relevance. An efficient integrator of the resulting dynamical equations which assures a satisfactory sampling of the phase space is based on the  decoupling of the harmonic spring term of the Hamiltonian by transforming the primitive coordinates into the normal mode coordinates: $V_{harmonic}({\bf X}_{I})=\frac{1}{2}M_{I}\omega_{P}^{2}{\bf X}_{I}^{T}A{\bf X}_{I}$, where $A$ is the matrix that couples the coordinates of different beads.
Once the matrix is diagonalised then the eigenvectors are used to represent the Hamiltonian in normal mode coordinates: $H_{nm}=\sum_{i=1}^{P} \left[ \frac{p_{i}^{2}}{2m_{i}^{'}} + \frac{1}{2}m\omega_{P}^{2} \lambda_{i}(x_{i}^{'})^{2} +  \frac{1}{P}V(x_{i}({\bf X^{'}}))\right]$, with $\lambda_{i}$ the $i$-$th$ eigenvalues of the diagonalized matrix.
Here for simplicity we have reported the one-particle Hamiltonian only. 
The equations of motion can then be written in terms of normal mode variables and the different choice of the fictitious mass in the equations leads to different PIMD algorithms~\cite{witt},  
although the methods differ considerably from the conceptual point of view. Ring Polymer 
Molecular Dynamics (RPMD)~\cite{rpmd} gives an approximation to Kubo-transformed correlation 
functions by using classical 
MD trajectories in the extended phase space of polymer rings. RPMD, however, suffers from the so-called
``resonance-problem''~\cite{witt, man} which causes a spurious splitting of the stretching peak in the IR spectrum. 
Thermostated RPMD (TRPMD)~\cite{rossi} is an improvement over the conventional RPMD method, where the spurious splitting
is removed by coupling the internal modes of the ring polymer to a thermostat. Centroid Molecular Dynamics (CMD)~\cite{caovoth} is based on the 
evolution of the centroid of the ring polymer on the potential energy surface created by the internal modes of the ring. CMD will be discussed in detail in Section~\ref{cmdsec}. Alternative methods for treating quantum dynamics, outside the realm of path integral techniques, are those such as Linearized semi-classical
initial value representation (LSC-IVR) method~\cite{lscivr1, lscivr2, lscivr3}; it uses classical MD trajectories and adds quantum effects
using the initial value representations (IVR)~\cite{ivr1, ivr2} of semi-classical theory~\cite{semiclassical}. This approach, however, does not 
conserve the quantum Boltzmann distribution. Furthermore, Althorpe and co-workers~\cite{althorpe, althorpe1} have recently proposed a method
called ``Matsubara dynamics'' which originates from a single change in the derivation of LSC-IVR method and generates classical dynamics and conserves the quantum Boltzmann distribution. They have also given the error terms in the propagator between exact quantum dynamics and CMD (as well as RPMD and TRPMD).
Within the context of Grand Canonical Adaptive Resolution some PIMD approaches have already been discussed (see Ref.~\cite{jcppi}), thus here we will discuss the implementation and application of Centroid Molecular Dynamics in GC-AdResS. 
\subsubsection{Centroid Molecular Dynamics}~\label{cmdsec}
A centroid is a quasi-classical object that is defined as an 
average over all the beads in a ring polymer as described before: $x_{c} = \frac{1}{P} \sum_{i=1}^{P} x_{i}, p_{c} = \frac{1}{P} \sum_{i=1}^{P} p_{i}$, and the resulting dynamics is named Centroid Molecular Dynamics (CMD)  ~\cite{caovoth}.
In this context, the normal mode transformation reported before is the optimal choice of CMD simulations; in fact the centroid separates out the first normal mode coordinate from the other modes. The evolution of the 
centroid is then governed by the following equations:
\begin{equation}
\dot{x}_{c}=\frac{p_{c}}{m}
\end{equation}
and,
\begin{equation}
m_{c}\ddot{x_{c}}=-\frac{\partial V_{o}(x_{c})}{\partial x_{c}} 
\end{equation}
where $m_{c}$ is the physical mass and $V_{o}$ is the potential of mean force generated by the dynamics of the non-centroid modes. 
The rigorous CMD procedure involves an accurate sampling of the phase space pertaining to the non-centroid modes at each position of the centroid. 
Such a procedure is computationally highly expensive, and thus one uses adiabatic decoupling to separate the fictitious motion of the non-centroid modes from 
the physical motion of the centroid. This version of Centroid Molecular Dynamics is called Adiabatic Centroid Molecular Dynamics (ACMD)~\cite{witt, pavese, muser}.
The adiabatic decoupling is achieved by reducing the masses of the non-centroid modes by a factor $\gamma^{2}$, 
where $0<\gamma^{2}<1$. The effect is that the motion of the centroid is slower compared to the the non-centroid modes, which implies
that the centroid moves on the potential of mean force generated ``on-the-fly" by rest of the modes. Thus, the choice of mass in CMD is:
\begin{equation}
m_{i}^{'}=\gamma^{2}m\lambda_{i}, m_{1}^{'}=m
\end{equation}
where $\gamma$ is the adiabaticity factor. 
There exists another formulation of ACMD, called partially ACMD (PACMD)~\cite{hone,man}, with the only difference that larger values of $\gamma$ are used in PACMD.
Due to a partial separation between the non-centroid and centroid modes, PACMD can be computationally less expensive than other PI-based approaches designed for the calculation of dynamic properties such as RPMD \cite{man}. It was shown 
in Ref.~\cite{hone} that the dynamical properties for liquid parahydrogen were similar with both ACMD and PACMD methods. In this work, we have implemented
PACMD in GC-AdResS and from now on, we will refer to PACMD as CMD. 
It must be reported that the vibrational spectra in CMD suffers from the curvature problem, due to which 
the stretching peak in the spectra is red-shifted and broadened as the temperature is lowered~\cite{witt}. It 
has been shown by Ivanov et al.~\cite{ivanov} that the curvature problem exists in CMD simulations of liquid water.
In this perspective this work must be evaluated for its technical significance regarding the computational implementation, i.e. the capability of reproducing conventional CMD result; the simulation carries the same physical limitation of conventional CMD results.
Paesani et.al.~\cite{medders} have recently shown that the effects of curvature problem have negligible effects when MB-pol potential energy 
surface is used in adiabatic CMD.  In such a scenario, CMD-GC-AdResS will not carry the current limitations, since the validity of GC-AdResS is independent of the specific molecular model.
\subsection{GC-AdResS}
The Grand Canonical Adaptive Resolution Simulation approach (GC-AdResS) is a multiscale technique that allows to couple different molecular models which describe the molecules in question at different levels of resolution. (see Fig.\ref{cmd}).
The simulation box is divided in three parts: (i) high resolution region, (ii) hybrid or transition region, (iii) coarse-grained region. In the current case the high resolution region is where molecules are described via the path integral approach and where the CMD technique is applied, instead the transition region is a technical filter which allows to pass from the PI representation to a coarse-grained representation. Finally in the coarse-grained region molecules are treated as generic classical spheres (without any quantum characteristic) interacting via a generic WCA potential (see Figure~\ref{cmd}).
It has been shown that the approach is, in general, theoretically well founded \cite{prx,njp} and numerically solid; moreover, in recent years, the method has been successfully extended to several approaches based on the PI representation of atoms \cite{prlado,jcppi}.
The technical implementation of CMD in GC-AdResS follows from the general implementation of PI representation in GC-AdResS as reported in Ref.\cite{jcppi}, however the capability of  GC-AdResS CMD to deliver correct results strongly relies on its capability to sample the correct phase space according to the CMD procedure (see also note \cite{ensem}); the aim of this paper is to show such an accuracy/efficiency. 
\subsection{Implementation of Centroid Molecular Dynamics  in GC-AdResS} 
\begin{figure}
  \centering
  \includegraphics[width=0.75\textwidth]{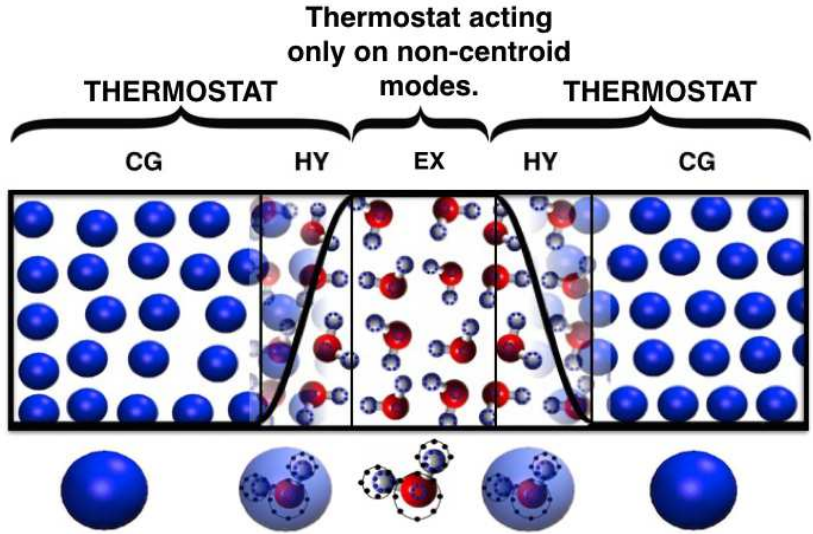}
  \caption{ Pictorial representation of the GC-AdResS scheme; CG indicates the coarse-grained region, HY the hybrid region where path-integral and coarse-grained forces are interpolated via a space-dependent, slowly varying, function $w(x)$ and EX (or PI) is the path-integral region (that is the region of interest). In the explicit path-integral
  subregion, the centroid mode is not subjected to any thermostat, while non-centroid modes move under the action of the thermostat.}
  \label{cmd}
\end{figure}

Since the path-integral polymer rings can be interpreted in terms of classical fictitious atoms (beads) with harmonic 
interaction between the adjacent beads, the standard equation of GC-AdResS can be used in a straightforward way. 
\begin{equation}
{\bf F}_{\alpha \beta}=w(X_\alpha)w(X_\beta){\bf
  F}_{\alpha\beta}^{PI}+[1-w(X_\alpha)w(X_\beta)]{\bf F}^{CG}_{\alpha\beta} 
\label{eqforce}
\end{equation}
where $\alpha$ and $\beta$ indicate two molecules, ${\bf F}^{PI}$ is the
force derived from the path-integral force field and  ${\bf F}^{CG}$ is the force derived from a generic coarse-grained potential, $X$ is the $x$ coordinate of the center of mass of the molecule and $w$ is an interpolating function which smoothly goes from $0$ to $1$ (or vice versa) in the interface region, ($\Delta$), where the lower resolution is slowly transformed (according to $w$) in the high resolution (or vice versa). 
This equation represents the coupling of two different regions characterized by different number of (effective) classical degrees of freedom \cite{note1}. A thermodynamic force, acting on the center of mass of each molecule in the transition region, is introduced in GC-AdResS to balance  the pressure difference between the coarse-grained and the explicit path-integral region \cite{jcpsimon,prl12,prx,njp} and it is numerically implemented via the following iterative procedure: 
\begin{equation}
F_{k+1}^{th}(x)=F_{k}^{th}(x) - \frac{M_{\alpha}}{[\rho_{ref}]^2\kappa}\nabla\rho_{k}(x)
\end{equation}
with $M_{\alpha}$ the mass of the molecule, $\kappa$ a constant which can be chosen in an appropriate way, $\rho_{o}$ is the target, average, density of reference and $\rho_{k}(x)$ is the molecular density at the $k$-th iteration as a function of the position in the transition/hybrid region.
The iteration converges when the density profile across the HY region becomes flat (within max $2-3\%$ of deviation from $\rho_{ref}$); such a force is very sensitive to thermodynamic conditions and numerical integrators, thus its implementation (and resulting accuracy) in the CMD version of GC-AdResS, despite we follow the same successful protocol of previous PI-CG-AdREsS calculations, is not a trivial numerical result. 
The dynamics of the non-centroid modes in CMD is artificial and is carried out in order to sample $V_{o}(x_{c})$. Such a process requires a canonical sampling 
over the non-centroid modes, which is achieved by  coupling the internal modes to a thermostat for rapid equilibration~\cite{witt, muser}. Since the dynamics of the centroid mode
is Newtonian, there are no thermostats attached to the centroid.
In the context of AdResS, this would simply translate to having a thermostat in the 
coarse-grained and hybrid regions, while in explicit region, no thermostats are attached to the centroid mode, while non-centroid modes
move under the action of the thermostat. Figure~\ref{cmd} shows the GC-AdResS CMD system with the application of a thermostat in different regions.
\section{Calculation of Quantum Dynamical properties in GC-AdResS}
The Kubo transformed quantum time correlation between two operators $\hat{A}$ and $\hat{B}$ is approximated by~\cite{muser}:
\begin{equation}
C_{AB}(t) = \frac{1}{Q} \int \frac{dx_{c}dp_{c}}{2\pi\hbar} A(x_{c}(0)) B(x_{c}(t)) e^{-\beta H_{c}}
\end{equation}
where $Q$ is the canonical partition function and $H_{c} = p_{c}^{2}/2m + V_{o}(x_{c})$ is the Hamiltonian governing the evolution of 
centroids on the potential $V_{o}(x_{c})$ created by the internal modes. 
The direct extension of the above formula to the case of an open boundary system/grand-canonical-like ensemble is:
\begin{equation}
C_{AB}(t)=\frac{1}{Q_{GC}}\sum_{N}\frac{1}{N!}\int \frac{dx_{c}^{N} dp_{c}^{N}}{2\pi\hbar} A(x_{c}^{N}(0)) B(x_{c}^{N}(t)) e^{-\beta \left[H_{c}^{N}(x_{c}^{N}, p_{c}^{N}) - \mu N\right]}
\end{equation}
where $Q_{GC}$ is the grand-canonical partition function, $\mu$ is the chemical potential, $N$ is the number of path 
centroids (which is now a variable number in the system) and $H_{c}^{N}$ is the Hamiltonian governing the evolution of centroids with $N$,
instantaneous number of path centroids. The number of centroids $N^{'}$ at time $t=t^{'}>0$ is likely to be different from the number of centroids $N$ at time $t=0$.
We use the correspondence between the Bergmann and Lebowitz model \cite{bl1,bl2,lebshim} and GC-AdResS to interpret the quantity $B(x_{c}^{N}(t))$ in this context. From such a correspondence one concludes that there exists a Liouvillian operator $iL^{M}$, which evolves the system from the configuration in phase space ${\bf X}_{c}^{N}(0)$ to the configuration in phase space ${\bf X}_{c}^{N^{'}}(t)$ (thus from $N$ to $N^{'}$ molecules) \cite{njp,pre16}. 
From the numerical point of view, we use the same procedure as used in Ref.~\cite{njp} to calculate the equilibrium time correlation functions, which 
is based on the definition of reservoir in  the Bergmann and Lebowitz model. According to this model, if a molecule leaves the system and enters
in the reservoir, it does not retain its microscopic identity. Thus if a molecule present time $t$ moves into a reservoir at time $t^{'}\le t$, then it does not contribute to the correlation after time $t$ (see note \cite{loc} and Ref.\cite{pre16}). 

\section{Results and Discussion}

\subsection{Low-temperature Parahydrogen: Technical Details}
All simulations in this work have been performed in home modified GROMACS MD package~\cite{gromacs} and the thermodynamic
force has been calculated using VOTCA simulation package~\cite{votca}. 
We have performed simulations of parahydrogen at two different temperatures: 14K and 25K. The number of parahydrogen
molecules in the system is 3000, and the box dimensions are chosen to reproduce the experimental density~\cite{paradens}: $\rho=26.2 cm^{3}/mol$ 
at 14K and $\rho=31.2 cm^{3}/mol$ at 25K. This corresponds to box dimensions $90\times38\times38$ $\angstrom^3$ at temperature 14K and 
$90\times41\times41$ $\angstrom^3$ at temperature 25K. In AdResS simulations, the resolution of the molecules changes along x-axis, as depicted in Figure~\ref{cmd}.
The size of the quantum and transition region is 20 $\angstrom$. 
The intermolecular interaction is described as in previous work \cite{para1, para2} by the (spherical) Silvera-Goldman potential \cite{silvgold} and the cut-off used is 9 $\angstrom$.
In the coarse-grained region, we have used a generic WCA potential given by:
\begin{equation}
U(r) = 4\epsilon\bigg[\bigg(\frac{\sigma}{r}\bigg)^{12} - \bigg(\frac{\sigma}{r}\bigg)^{6}\bigg]  + \epsilon,            r \leq 2^{1/6}\sigma
\end{equation}
For parahydrogen at 14K, the parameters $\sigma$ and $\epsilon$ are 0.30 nm and 0.90 kJ/mol respectively, and 
for the system at 25K, $\sigma$ and $\epsilon$ are 0.30 nm and 0.80 kJ/mol respectively
We have used $P=48$ beads for $T=14K$ and $P=32$ beads for $T=25K$. These values give 
converged results for low temperature parahydrogen~\cite{beads, hone}.
An adiabaticity parameter of $\gamma^{2}=1/P$~\cite{hone}
is used and a time step of 0.25 $fs$ is found to be optimal for the corresponding adiabaticity parameter. A 200 ps long PIMD simulation is performed and along the trajectory, 
configurations are stored every 4 ps to perform CMD simulations. Thus, a total of 50 trajectories each of length 10 ps are generated. 
For the first 4 ps, we keep all the modes under the action of thermostat, in order to adjust the velocities as masses are different 
in PIMD and CMD methods. We have strictly followed the procedure reported in the work of Perez, Tuckerman and M\"{u}ser \cite{muser}. After this initial equilibration run, centroid mode is not kept under the action of the thermostat while  
non-centroid modes are thermostated.  We use the same strategy for AdResS simulations, 
where a 200 ps long fully thermostated GC-AdResS PIMD simulation is performed, and 50 initial configurations are taken along 
this trajectory to perform GC-AdResS CMD simulations. For the first 4 ps, the thermostat acts in the explicit as well as the 
hybrid and coarse-grained regions. After the short equilibration run, the centroid modes are not coupled to a thermostat in the explicit region, while 
the hybrid and coarse-grained region are kept under the action of the thermostat. The equilibrium properties are calculated in the 
explicit region in the last 6 ps, i.e. excluding the equilibration run.  The velocity auto correlation function is calculated up to 2 ps 
by averaging over the 50 trajectories. The diffusion coefficient is obtained from the time integral of the velocity auto-correlation function:
\begin{equation}
D=\frac{1}{3} \int_{0}^{\infty} C_{vv}(t) dt
\end{equation}

\subsection{Low-temperature Parahydrogen: Results}
\begin{figure}
\centering
\subfigure[14K]{\label{fig:1a}\includegraphics[width=2.7in]{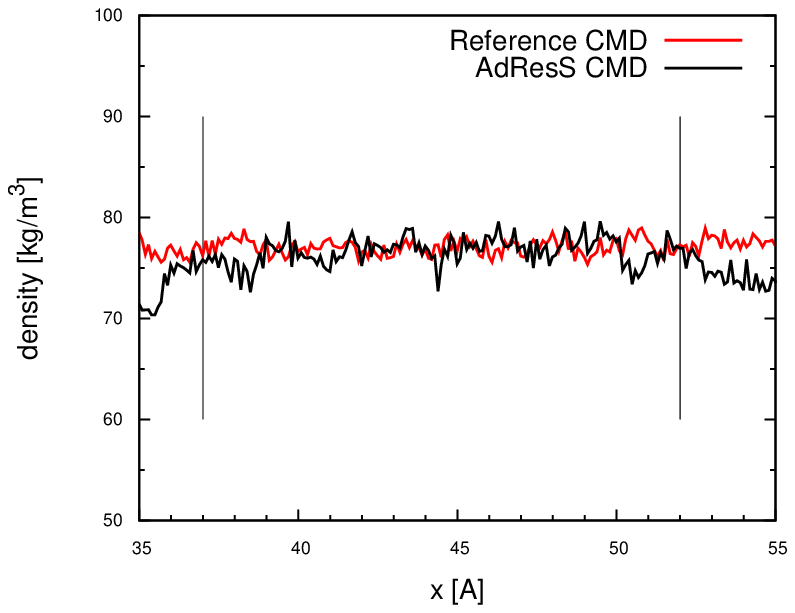}}
\subfigure[25K]{\label{fig:1b}\includegraphics[width=2.7in]{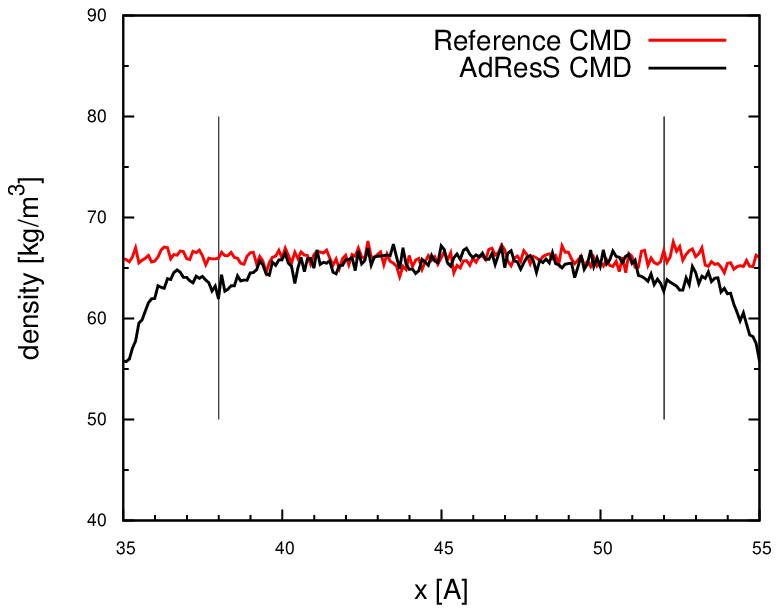}}
\caption{Centroid density in the explicit path-integral region in reference CMD and AdResS CMD simulations for liquid parahydrogen. All the properties are calculated in the region enclosed between the vertical lines using the rigorous GC-AdResS protocol.}
\label{denspar}
\end{figure}

\begin{figure}
\centering
\subfigure[14K]{\label{fig:2a}\includegraphics[width=2.7in]{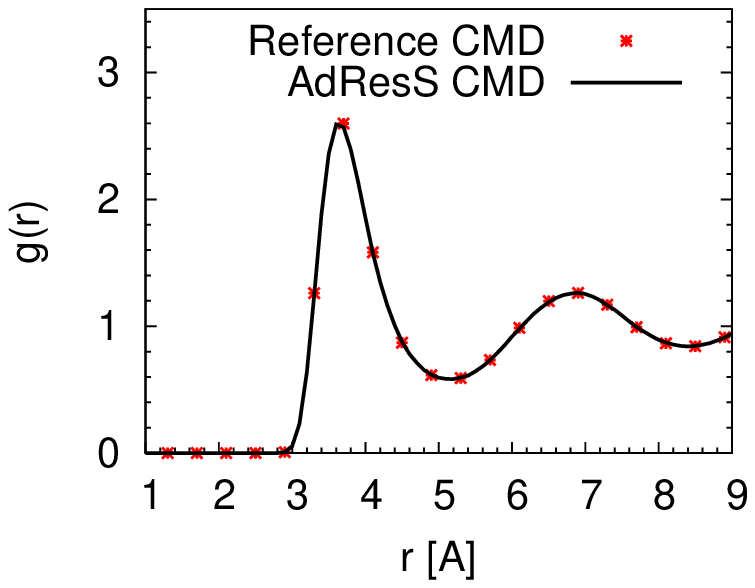}}
\subfigure[25K]{\label{fig:2b}\includegraphics[width=2.7in]{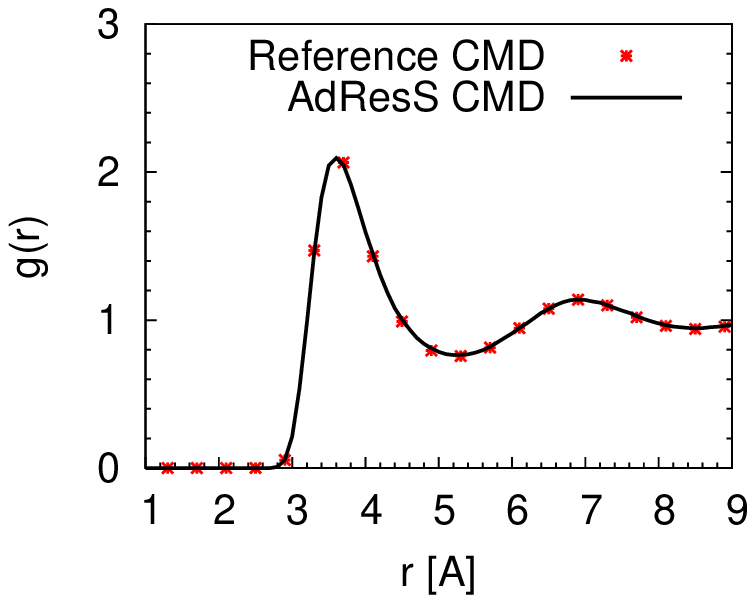}}
\caption{Centroid Radial distribution functions calculated in the explicit region in AdResS CMD and an equivalent region in the reference CMD simulations for liquid parahydrogen.}
\label{rdfpar}
\end{figure}

\begin{figure}
\centering
\subfigure[14K]{\label{fig:3a}\includegraphics[width=2.7in]{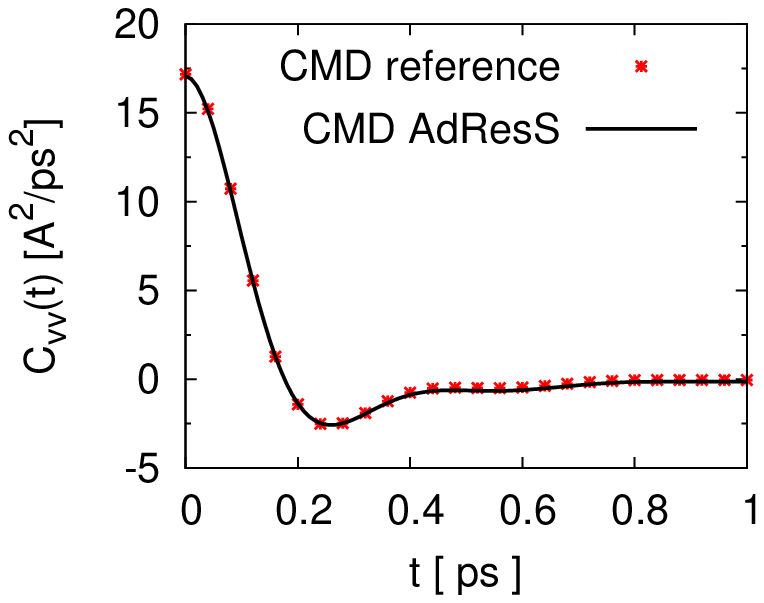}}
\subfigure[25K]{\label{fig:3b}\includegraphics[width=2.7in]{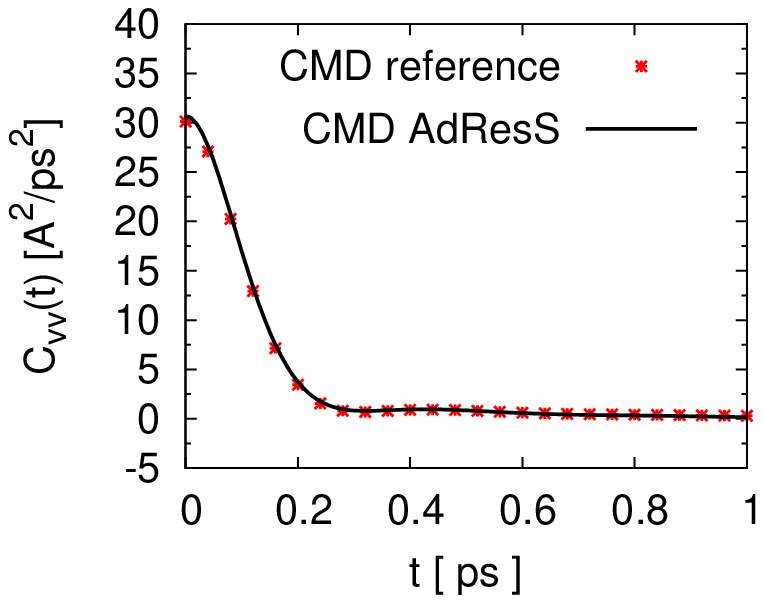}}
\caption{Velocity auto-correlation function calculated in the explicit region in AdResS CMD and an equivalent region in the reference CMD simulations for liquid parahydrogen.}
\label{vacpar}
\end{figure}

Figure~\ref{denspar} shows the centroid density for low-temperature parahydrogen in the explicit path-integral subregion. 
The agreement between the reference results and AdResS results is highly satisfactory. In a rigorous application of the GC-AdResS protocol, a part of the explicit region, in contact with the transition/hybrid region is considered as a buffer region and it is included in the transition region; however if this is not done an error of, at worst, $10\%$ should be considered or a more accurate (but more expensive) thermodynamic force shall be calculated. Following the prescription above, in this work all the properties are calculated in the explicit path-integral subregion 
excluding the border areas (as shown by the two vertical lines in Figure~\ref{denspar}). Figure~\ref{rdfpar} shows the centroid radial distribution 
functions for low temperature parahydrogen calculated in the explicit region in AdResS CMD and an equivalent region in reference CMD simulations. 
The results are highly satisfactory. It should be noted that centroid RDF's are not same as the quantum (bead-bead) RDF's and a deconvolution 
procedure~\cite{convu} is used to convert centroid RDF's to the actual quantum RDF's. However, it is an important numerical quantity to show that the explicit path-integral region in  AdResS reproduces detailed structural properties in a proper way. Figure~\ref{vacpar} shows the velocity auto-correlation function for low-temperature parahydrogen calculated in the explicit region 
in AdResS CMD and an equivalent region in reference CMD simulations. It can be seen that AdResS CMD results and the reference CMD results agree in a rather satisfactory way.
This has also been verified by comparing the local diffusion constant calculated by integrating over the velocity auto-correlation functions, which can be seen 
in Table~\ref{table1}.

\begin{table}
\centering
\caption{Local diffusion constant $D$ ($\angstrom^{2} ps^{-1}$)  for liquid parahydrogen.}
\label{table1}
\begin{tabular}{ccc}
\hline \hline
Temperature & Reference CMD & AdResS CMD \\
\hline
14K & 0.37  & 0.33 \\
25K & 1.37  & 1.36 \\
\hline \hline
\end{tabular}
\end{table}

\begin{figure}
  \centering
  \includegraphics[width=2.7in]{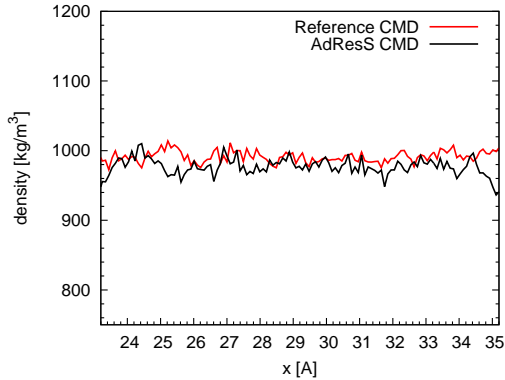}
  \caption{Centroid density in the explicit path-integral region in reference CMD and AdResS CMD simulations for liquid water.}
  \label{dens}
\end{figure}

\begin{figure}
  \centering
  \includegraphics[width=2.7in]{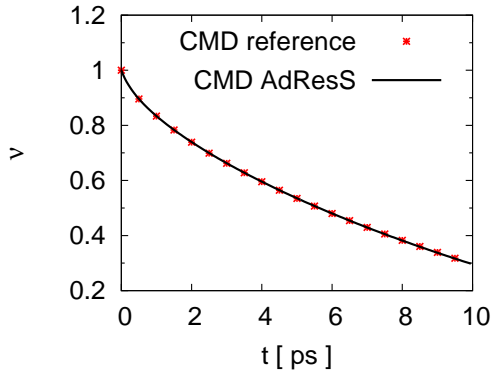}
  \caption{Number of molecules that remain within the explicit path-integral region $\nu$ as a function of time. This quantity is calculated in reference CMD and AdResS CMD simulations.}
  \label{drift}
\end{figure}

\begin{figure}
\centering
\subfigure[H-H g(r)]{\label{fig:4a}\includegraphics[width=2.7in]{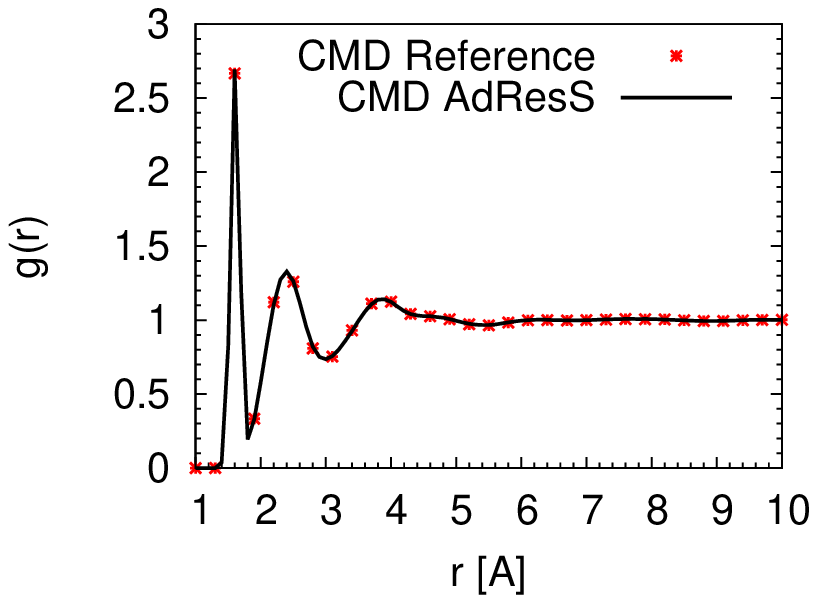}}
\subfigure[O-H g(r)]{\label{fig:4b}\includegraphics[width=2.7in]{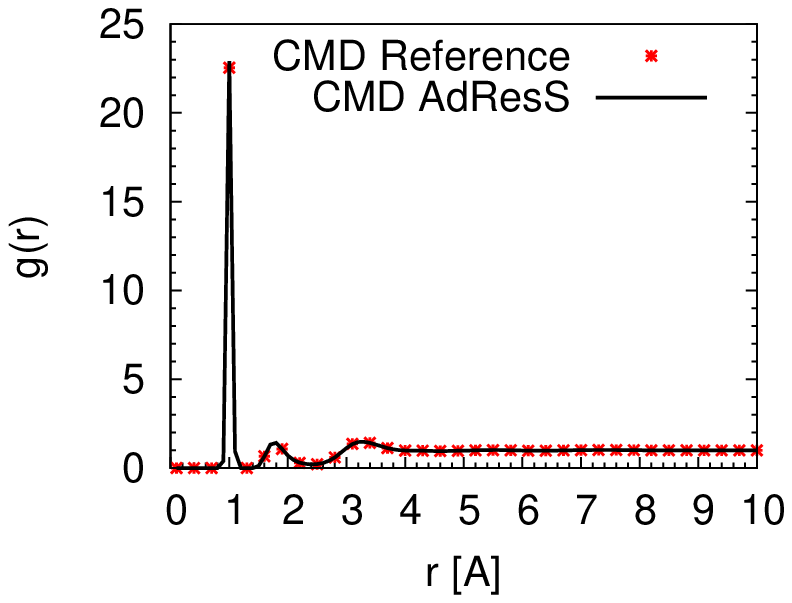}}
\subfigure[O-O g(r)]{\label{fig:5a}\includegraphics[width=2.7in]{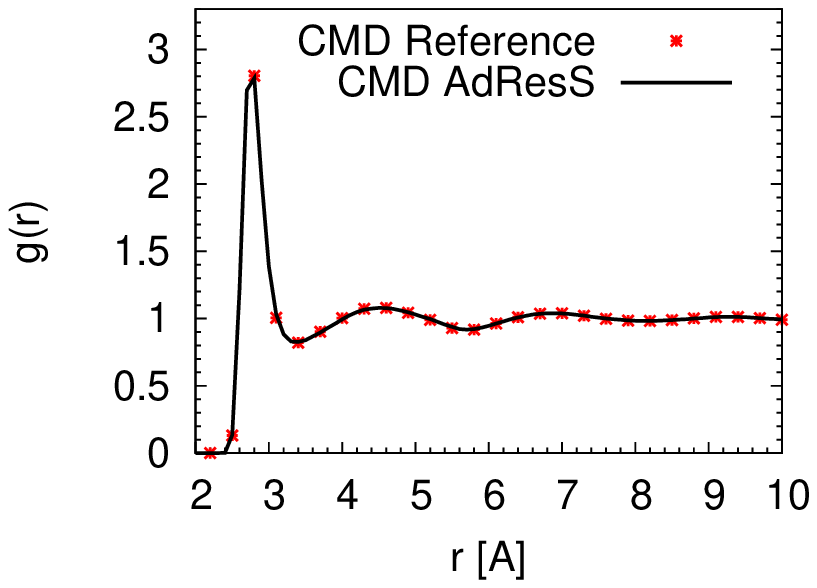}}
\caption{Centroid radial distribution functions for liquid water calculated in explicit region of AdResS CMD and an equivalent subregion in the reference CMD simulations.}
\label{rdf}
\end{figure}

\begin{figure}
\centering
\subfigure[Velocity autocorrelation function]{\label{fig:6a}\includegraphics[width=2.7in]{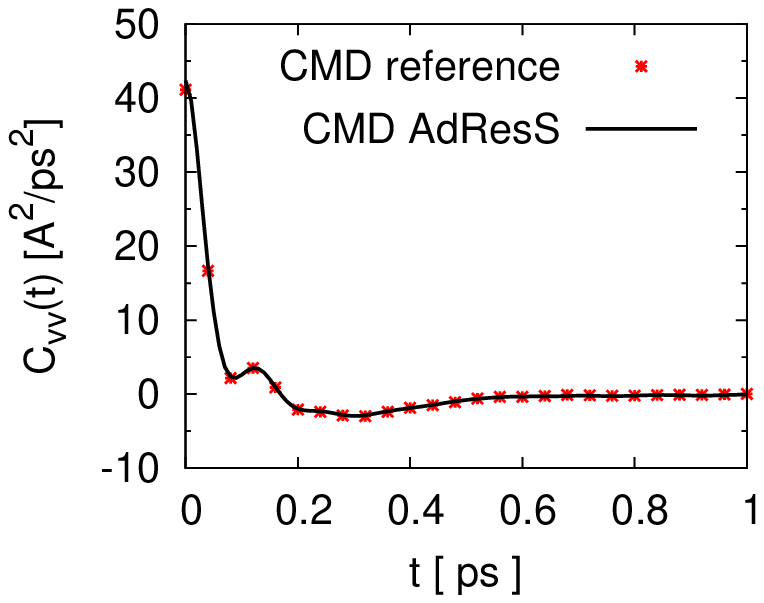}}
\subfigure[First order orientational correlation function]{\label{fig:5b}\includegraphics[width=2.7in]{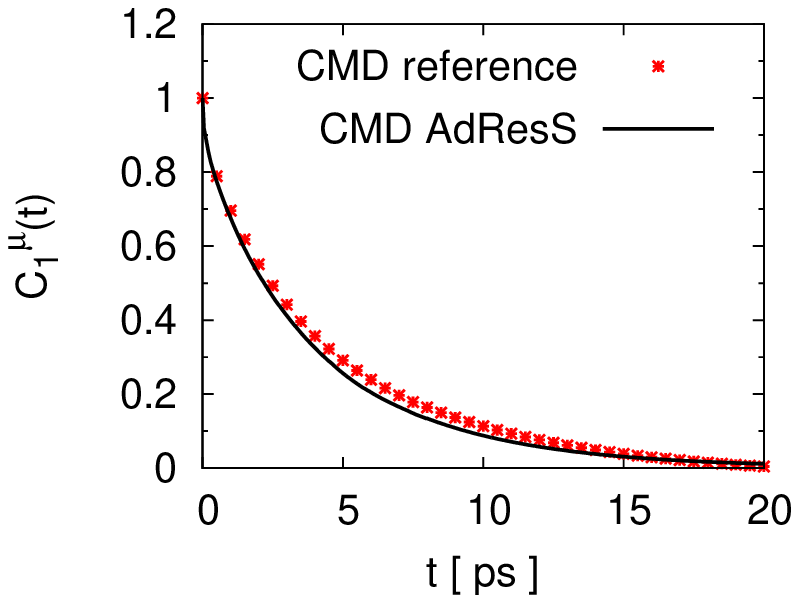}}
\subfigure[Second order orientational correlation function]{\label{fig:7a}\includegraphics[width=2.7in]{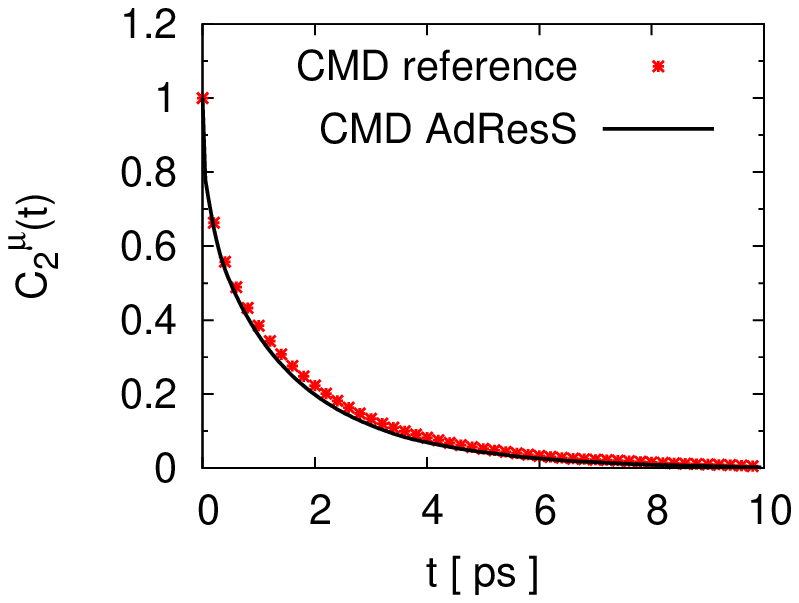}}
\caption{Equilibrium time Correlation Functions for liquid water calculated in explicit region of AdResS CMD and an equivalent subregion in the reference CMD simulations. For the 
first order correlation function, an exponential tail has been fitted beyond 10 ps. }
\label{vac}
\end{figure}

\begin{table}
\centering
\caption{Local diffusion constant and $l^{th}$ order relaxation times for liquid water calculated in explicit region of AdResS CMD and an equivalent subregion in the reference CMD simulations.}
\label{table2}
\begin{tabular}{ccc}
\hline \hline
Parameter & Reference CMD & AdResS CMD \\
\hline
$D$ ($\angstrom^{2} ps^{-1}$)  & 0.32 & 0.32  \\
$\tau_{1}^{\mu}$ ($ps$) & 4.0  & 3.7  \\
$\tau_{2}^{\mu}$ ($ps$) & 1.3  & 1.2   \\
\hline \hline
\end{tabular}
\end{table}
\begin{figure}
  \centering
  \includegraphics[width=2.7in]{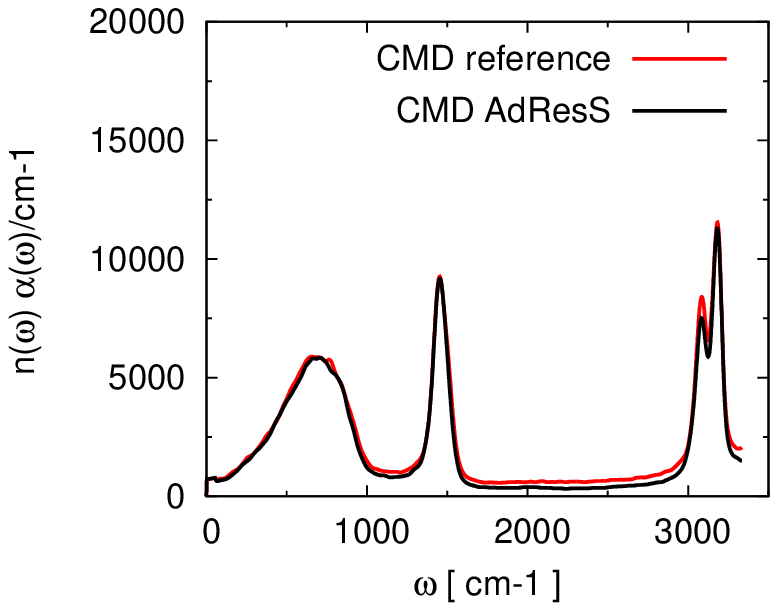}
  \caption{Infrared spectrum for liquid water at 298K calculated in explicit region of AdResS CMD and an equivalent subregion in the reference CMD simulations.}
  \label{spectrum}
\end{figure}
\subsection{Liquid Water: Technical Details}
We use the q-SPC/Fw water model~\cite{paesani} for performing CMD simulations. 
The number of water molecules in system are 1320, and the box dimensions are $58\times26\times26$ $\angstrom^3$, corresponding to a density $990$ $kg m^{-3}$.  
The size of the quantum subregion treated in this work is 12 $\angstrom$ and the size of the transition region is 24 $\angstrom$. The remaining system contains 
coarse-grained particles, which interact via generic WCA potential.
The parameters $\sigma$ and $\epsilon$ in the current simulations are 0.30 nm and 0.65 kJ/mol respectively.
The number of imaginary time slices is taken to be $P=32$ and an adiabaticity parameter of $\gamma^{2}=P^{-(P+1)/(P-1)}$~\cite{man}
is used, and a time step of 0.1 $fs$ is found to be sufficient for this adiabaticity parameter.  
Reaction field method is used to compute the electrostatic properties with dielectric 
constant for water = $80$. The cut-off for both Van Der Waals and electrostatic interactions is 9 $\angstrom$.
We generate a 200 ps long PIMD trajectory, where the configurations are stored after every 8 ps.
We initiate 25 ps long CMD trajectories from the saved configurations. 
All the modes are coupled to a thermostat for first 5 ps. After this initial warm up run, we decouple the centroid mode 
from the thermostat while non-centroid modes move under the action of thermostat.  
In the CMD AdResS simulations, this translates to having a thermostat coupled to the whole system for the first 5 ps, 
following which the thermostat is coupled to all the modes in coarse-grained and hybrid regions and only non-centroid 
modes in the explicit path-integral region.  
The dynamic properties are calculated in the
explicit region in the last 20 ps, i.e. excluding the equilibration run.  The correlation functions are calculated 
up to 10 ps by averaging over the 25 trajectories. 
We calculate the $l^{th}$ order orientational correlation functions of the type:
\begin{equation}
C_{l}(t)=\langle P_{l} [\vect e(0) \cdot \vect e(t)] \rangle,
\end{equation}
where $P_{l}$ is the Legendre polynomial of order $l$, and $\vect e$ is a unit vector 
that is chosen along one of the three principle inertial axes of the water molecule.
The $l^{th}$ order relaxation times are obtained from the time integrals of the 
corresponding orientational correlation functions:
\begin{equation}
\tau_{l}^{n}=\int_{0}^{\infty} C_{l}^{n} dt
\end{equation}  
Since the orientational correlation functions are calculated for 10 ps, an exponential tail was fitted to the correlation functions 
for computing the integral. 
We also calculate the infrared absorption coefficient $\alpha(\omega)$ using the following relation~\cite{paesani}:
\begin{equation}
\alpha (\omega) = \left[ \frac{4\pi^{2} \omega}{3V\hbar cn(\omega)} \right] (1-e^{-\beta \hbar \omega}) \frac{1}{2\pi} \times \int_{-\infty}^{+\infty} e^{-i\omega t} \langle M(0)M(t) \rangle dt, 
\end{equation}
where $\langle M(0)M(t) \rangle$ is the total dipole moment auto-correlation function, $c$ is the speed of light, $V$ is the volume of the box and $n(\omega)$ is the refractive index of the system at frequency $\omega$.
\subsection{Liquid Water: Results}
Figure~\ref{dens} shows the centroid density for liquid water in the explicit path-integral subregion. 
The agreement between the reference results and AdResS results is highly satisfactory.
Figure~\ref{drift} shows the number of molecules that remain within the explicit path-integral region $\nu$ as 
a function of time. This is calculated as following: we label all the molecules in the trajectory at time `0' and calculate how many 
of those labeled molecules are present at time `$t$'. This is an important quantity that describes the movement of the molecules 
in and out of the explicit path-integral region. This quantity is calculated in AdResS CMD and reference CMD simulations. It can be 
seen that the two curves overlap. This result confirms, once again, that AdResS subregion has the same average dynamical behaviour as the reference CMD subregion, and indirectly shows the Grand Canonical-like character of GC-AdResS. Figure~\ref{rdf} shows the centroid RDF's calculated in the explicit region in AdResS CMD and 
an equivalent region in reference CMD simulations. The results are highly satisfactory. Figure~\ref{vac} shows the velocity auto-correlation 
function, first and second order orientational correlation functions (by defining the unit vector along the direction of molecular dipole moment) 
calculated in the explicit region in AdResS CMD and an equivalent region in reference CMD simulations and Table~\ref{table2} reports the 
the local diffusion constant ($D$ ($\angstrom^{2} ps^{-1}$)) and $l^{th}$ order relaxation times ($\tau_{1}^{\mu}$ ($ps$) and $\tau_{2}^{\mu}$ ($ps$)).
It can be seen that the local diffusion constant is same in both AdResS CMD and reference CMD results, while there is some discrepancy in the 
$1^{st}$ order relaxation time. The difference is not significant, however it must be reported. Figure~\ref{spectrum} shows the infrared spectrum calculated in the path-integral subregion in AdResS CMD and reference CMD simulations. The agreement is remarkable, and strongly supports the numerical and conceptual solidity of the method since the spectrum is a quantity of primary importance also from an experimental point of view. In general, it should be pointed out that the current GC-AdResS simulations 
are not performed under optimal conditions, i.e. a very large reservoir and (ideally) a relatively small hybrid region. The computational set up employed in this work represents a ``worst case scenario'' that tests the technical frontiers of the method; it is natural to expect that when theoretical conditions are fully met then the level of accuracy can only rise (as proven recently \cite{njp}).
However, already under non optimal conditions the results are highly satisfactory. 
\section{Conclusions}
We have reported the implementation and testing application of CMD in the open boundary, Grand Canonical-like Adaptive Resolution Simulation technique. We have studied two test systems: (a) liquid parahydrogen at low temperature and (b) liquid water at ambient conditions. Structural and dynamical properties were calculated and compared with reference full CMD calculations, the results show a highly satisfactory agreement. The GC-AdResS set up can be also employed as a tool of analysis by systematically increasing/decreasing the quantum region and control whether some properties change when compared to the calculations of a full CMD system. This approach would allow for a determination of the essential degrees of freedom required for a certain property. In fact the reservoir is very generic and its only physical contribution is at macroscopic/thermodynamic level, thus as a matter of fact all the necessary degrees of freedom are exclusively those of the quantum region.
For classical systems this kind of approach has been already used to determine the locality of the hydrogen bonding network for water around large hydrophobic solutes \cite{jcpcov}. Interestingly, in  PI studies of systems as those in Ref.\cite{jcpcov}, one should add the effects of the quantum description to the intrinsic classical locality/non-locality described by the classical GC-AdResS. This implies that the use of GC-AdResS with PI methods would allow for the understanding, at a very basic/essential level, of the relevant principles behind the difference between classical and quantum results. In this paper we have shown that GC-AdResS CMD is technically robust and thus we can confidently foresee in future applications an analysis aimed at identifying relevant degrees of freedom at the level described by the PI approach. 
\section*{Acknowledgment}
This research has been funded by Deutsche Forschungsgemeinschaft (DFG) through grants CRC 1114 and DE 1140/7-1. We thank Shelby C. Straight and Prof. Fransesco Paesani at University of California, San Diego for providing us the scripts to calculate infrared spectrum. We thank Matej Praprotnik for a critical reading of the paper.


\begin{thebibliography}{24}
\expandafter\ifx\csname natexlab\endcsname\relax\def\natexlab#1{#1}\fi
\expandafter\ifx\csname url\endcsname\relax
  \def\url#1{\texttt{#1}}\fi
\expandafter\ifx\csname urlprefix\endcsname\relax\def\urlprefix{URL }\fi
\bibitem{fey-hibbs}
R.P.Feynman and A.R.Hibbs, {\it Quantum Mechanics and Path Integrals}, McGraw-Hill, Inc. 1965.
\bibitem{tuckerbook}
M.E. Tuckerman, {\it Statistical Mechanics: Theory and Molecular Simulation}, Oxford University Press, New York 2010.
\bibitem{man1}
M. Ceriotti, M. Parrinello, T.E. Markland and D.E. Manolopoulos, J. Chem. Phys. {\bf 133} 124104 (2010).
\bibitem{piglet2}
M. Ceriotti and D. E. Manolopoulos, Phys. Rev. Lett. {\bf 109}, 10064 (2012)
\bibitem{taka1}
S.Jang, G.Voth,  J. Chem. Phys. {\bf 115}, 7832 (2001).
\bibitem{taka2}
A.Perez and M.E.Tuckerman, J.Chem.Phys. {\bf 135}, 064104 (2011).
\bibitem{man3}
T.E. Markland and D.E. Manolopoulos, J. Chem. Phys. {\bf 129}, 024105 (2008).
\bibitem{cont2}
T. E. Markland and D. E. Manolopoulos, Chem.Phys.Lett. {\bf 464}, 256 (2008).
\bibitem{cont3}
G. S. Fanourgakis, T. E. Markland and D. E. Manolopoulos, J.Chem.Phys. {\bf 131}, 094102 (2009).
\bibitem{jcp1}
M.Praprotnik, L.Delle Site, and K.Kremer, J.Chem.Phys. {\bf 123}, 224106 (2005)
\bibitem{annurev}
M.Praprotnik, L.Delle Site, and K.Kremer, Annu.Rev.Phys.Chem. {\bf 59}, 545 (2008) 
\bibitem{ensing}
B.Ensing, S.O. Nielsen, P.B. Moore, M.L. Klein, and M.Parrinello, J.Chem.Th.Comp. {\bf 3}, 1100 (2007).
\bibitem{matej}
R. Delgado-Buscalioni, J. Sabli and M. Praprotnik, Eur. Phys. J. Special Topics {\bf 224}, 2509-2510 (2015).
\bibitem{hadress}
R.Potestio, S.Fritsch, P.Espanol, R.Delgado-Buscalioni, K.Kremer, R. Everaers and D.Donadio, Phys.Rev.Lett. {\bf 110}, 108301 (2013).
\bibitem{truh}
A.Heyden and D.G.Truhlar, J.Chem.Th.Comp. {\bf 4}, 217 (2008).
\bibitem{pande}
J.A. Wagoner and V. Pandey, J. Chem. Phys., {\bf 139}, 234114 (2013).
\bibitem{prx}
H.Wang, C.Hartmann, C.Sch\"{u}tte and L.Delle Site, Phys.Rev.X, {\bf 3}, 011018 (2013).
\bibitem{jcpchem}
A.Agarwal, H.Wang, C.Sch\"{u}tte and L.Delle Site, J.Chem.Phys. {\bf 141}, 034102 (2014)
\bibitem{njp}
A. Agarwal, J. Zhu, H. Wang  and L. Delle Site, New Jour.Phys., {\bf 17}, 083042 (2015).
\bibitem{jcppi}
A.Agarwal and L.Delle Site, J.Chem.Phys. {\bf 143}, 094102 (2015).
\bibitem{ensem}
In general GC-AdResS reproduces the same distribution of an open subsystem (of a large system) which exchanges heat and matter with the rest of the system. GC-AdResS is a Grand Ensemble, that is the ``reservoir'' does not need to have an infinite number of particles. However in previous work \cite{prx,njp} we have shown that, for a reasonable range of sizes of the coarse-grained and high resolution region usually employed in the GC-AdResS simulations, corrections due to the finiteness of the reservoir are not needed, and the distribution is essentially (effectively) Grand Canonical. For GC-AdResS with CMD (or equivalently for PIMD or RPMD) applies the same principle, since we use the isomorphism of the ``quantum statistical measure'' via the path integral approach and we reproduce the distribution of a subregion of the canonical full CMD system, then GC-AdResS reproduces the correct ``quantum distribution'' of an open boundary system (Grand Canonical distribution).
\bibitem{rpmd}
I.R. Craig and D.E. Manolopoulos, J. Chem. Phys., {\bf 121}, 3368 (2004).
\bibitem{manrev}
S. Habershon, D.E. Manolopoulos, T.E. Markland, and T.F. Miller III, Annu. Rev. Phys. Chem., {\bf 64}, 387 (2013)
\bibitem{hone}
T.D. Hone, P.J. Rossky and G.A. Voth, J. Chem. Phys., {\bf 124}, 154103 (2006).
\bibitem{expl}
RPMD employs a fictitious temperature function of the number of beads used, thus in RPMD the calculation of the thermodynamic force, necessary to properly run GC-AdResS (equilibration stage), must be done for each specific system (i.e. number of beads used). Instead in PIMD or CMD the thermodynamic force of an atomistic simulation (P=1) is sufficiently accurate for any system (i.e. number of beads) one chooses. For this reason, since the equilibration force done with $P=1$ is much cheaper than that done with $P=32$, CMD may be technically more convenient than RPMD. This aspect has been shown in detail, at numerically level, in our previous work \cite{jcppi}.
\bibitem{caovoth}
J. Cao and G.A. Voth, J. Chem. Phys., {\bf 99}, 10070 (1993)
\bibitem{witt}
A. Witt, S.D. Ivanov, M. Shiga, H. Forbert and D. Marx, J. Chem. Phys., {\bf 130}, 194510 (2009).
\bibitem{pavese}
M. Pavese, S. Jang, G.A. Voth, Parallel Computing, {\bf 26}, 1025-1041 (2000).
\bibitem{man}
S. Haberson, G.S. Fanourgakis and D.E. Manolopoulos, J. Chem. Phys., {\bf 129}, 074501 (2008).
\bibitem{muser}
A. Perez, M.E. Tuckerman, and M.H. Muser, J. Chem. Phys., {\bf 130}, 184105 (2009).  
\bibitem{prlado}
A.B.Poma and L.Delle Site, Phys.Rev.Lett. {\bf 104}, 250201 (2010).
\bibitem{note1}
It must be noted that in the current implementation of PIMD/RPMD/CMD in GROMACS, the  beads are retained in the coarse-grained region, and stay fixed relative to the path centroid. The algorithm works in 
the following way: In the coarse-grained region, the beads are treated as ghost particles, and do not interact with beads 
of other molecules. The forces are calculated between the centers-of-mass of the ring polymers in primitive coordinates, 
which are then redistributed over the beads of each ring polymer. Thus, the number of calculations for interatomic force are 
reduced in the coarse-grained region; which corresponds to 75\% of the computational cost. 
\bibitem{jcpsimon}
S.Poblete, M.Praprotnik, K.Kremer and L.Delle Site, J.Chem.Phys. {\bf 132}, 114101 (2010).
\bibitem{prl12}
S.Fritsch, S.Poblete, C.Junghans, G.Ciccotti, L.Delle Site and K.Kremer, Phys.Rev.Lett. {\bf 108}, 170602 (2012).
\bibitem{bl1}
J.L.Lebowitz and P.G.Bergmann, Phys.Rev. {\bf 99}, 578 (1955).
\bibitem{bl2}
J.L.Lebowitz and P.G.Bergmann, Annals of Physics, {\bf 1}, 1 (1957).
\bibitem{lebshim}
J.L.Lebowitz and A.Shimony, Phys.Rev. {\bf 128}, 1945 (1962).
\bibitem{loc}
GC-AdResS treats the high resolution region as an  open boundary system, thus the meaning of a time correlation function (or of correlation function in general) has a different physical interpretation compared to the equivalent quantity for closed systems. The time correlation function of an open boundary system is subject to the locality in space (i.e. dependence on the molecules present only in the high resolution region) and in time (i.e. dependence on the time scale of the residence of a molecule in the high resolution region). Thus the results of a open boundary system will coincide with those of a large full system if and only if the property of interest is local in space and in time. In this paper, from the technical point of view, we show the robustness of the method, by comparing the GC-AdResS results with those obtained by considering a subsystem of the whole system. Instead, if we compare GC-AdResS results with those obtained over the full system, then we can interpret this comparison only in terms of locality in space and time of a certain process. In time windows where GC-AdResS results reproduce those of a full CMD simulation (on the whole box), one has that, at least within that time frame, properties are local. When the ``deterioration'' of data, due the decrease of the number particles in the high resolution region as the time window becomes larger, becomes sizeable then the conclusion is that the property under observation is not local and needs a much larger ``bulk''; this, {\it per se} is a physical result. A detailed treatment of such concepts have been recently presented by one of us in Ref.\cite{pre16}.
\bibitem{pre16}
L. Delle Site, Phys. Rev. E. (2016) {\bf 93}, 022130 (2016).
\bibitem{gromacs}
S.Pronk, S.Pall, R.Schulz, P.Larsson, P.Bjelkmar, R.Apostolov, M.R.Shirts, J.C.Smith, P.M.Kasson, D.van der Spoel, B.Hess, E.Lindahl, Bioinformatics {\bf 29}, 845 (2013).
\bibitem{votca}
V. R\"{u}hle, C. Junghans, A. Lukyanov, K. Kremer, and D. Andrienko, J. Chem. Theory Comput. {\bf 5}, 3211 (2009).
\bibitem{beads}
T.F. Miller III and D.E. Manolopoulos, J. Chem. Phys. {\bf 122}, 184503 (2005).
\bibitem{paradens}
D. Scharf, G. J. Martyna, and M. L. Klein, Low. Temp. Phys., {\bf 19}, 365 (1993).
\bibitem{para1}
A.B.Poma and L.Delle Site, Phys.Chem.Chem.Phys. {\bf 13}, 10510 (2011).
\bibitem{para2}
R.Potestio and L.Delle Site, J.Chem.Phys. {\bf 136}, 054101 (2012).
\bibitem{silvgold}
I. F. Silvera and V. V. Goldman, J. Chem. Phys. {\bf 69(9)}, 4209 (1978).
\bibitem{convu}
N. Blinov and P.N. Roy, J. Chem. Phys. {\bf 120}, 3759 (2004).
\bibitem{paesani}
F. Paesani, W. Zhang, D.A. Case, T.E. Cheatham III, and G.A. Voth, J. Chem. Phys. {\bf 125}, 184507 (2006).
\bibitem{jcpcov}
B.P. Lambeth, C.Junghans, K.Kremer, C.Clementi, and L.Delle Site, J.Chem.Phys. {\bf 133}, 221101, (2010)
\bibitem{ivanov}
S.D. Ivanov, A. Witt, M. Shiga and D. Marx, J.Chem.Phys. {\bf 132}, 031101, (2010)
\bibitem{medders}
G.R. Medders and F. Paesani, J. Chem. Theory Comput. {\bf 11}, 1145−1154, (2015)
\bibitem{rossi}
M. Rossi, M. Ceriotti and D.E. Manolopoulos, J. Chem. Phys. {\bf 140}, 234116 (2014)
\bibitem{lscivr1}
H.Wang, X. Sun and W.H. Miller, J. Chem. Phys. {\bf 108}, 9726-9736 (1998)
\bibitem{lscivr2}
X. Sun, H. Wang and W.H. Miller, J. Chem. Phys. {\bf 109}, 7064-7074 (1998)
\bibitem{lscivr3}
J. Liu and W.H. Miller, J. Chem. Phys. {\bf 127}, 114506 (2007)
\bibitem{ivr1}
W.H. Miller, Proc. Natl. Acad. Sci. U.S.A. {\bf 102}, 6660 (2005)
\bibitem{ivr2}
W.H. Miller, J. Chem. Phys. {\bf 125}, 132305 (2006)
\bibitem{semiclassical}
W.H. Miller, Adv. Chem. Phys., {\bf 25}, 69 (1974)
\bibitem{althorpe}
T.J.H. Hele, M.J. Willatt, A. Muolo, and S.C. Althorpe J. Chem. Phys. {\bf 142}, 134103 (2015)
\bibitem{althorpe1}
T.J.H. Hele, M.J. Willatt, A. Muolo and S.C. Althorpe1, J. Chem. Phys. {\bf 142}, 191101 (2015)
\end{thebibliography}
\end{document}